\documentclass[12pt,nofootinbib]{revtex4}
\usepackage{amssymb}
\usepackage{amsmath, graphicx}
\usepackage{slashed}
\usepackage[unicode]{hyperref}
\usepackage{csquotes}
\usepackage{braket}
\usepackage{natbib}
\usepackage{array}
\usepackage{xcolor}

\begin{document}
\title{Dark matter production from evaporation of regular primordial black holes}
\author{Ngo Phuc Duc Loc}
\thanks{Email: locngo148@gmail.com}

\affiliation{Physics Division, National Center for Theoretical Sciences, National Taiwan University, Taipei 106319, Taiwan
}

\begin{abstract}
We point out that a simple redefinition of the regularizing parameter in regular black hole (RBH) metrics can preserve the self-similarity of the evaporation process. This implies that a RBH can evaporate completely, mirroring the behavior of its singular counterpart. Consequently, RBHs need not evolve into exotic, unverified remnant states such as horizonless compact objects or wormholes. We then provide a general framework to study dark matter (DM) production from evaporation of regular primordial black holes (RPBHs). As illustrative examples, we explicitly work out the cases of the Hayward metric and the Simpson-Visser metric. The formalism can be readily applied to other metrics. RPBH generally exhibits different Hawking temperature and horizon size compared to their singular counterpart, leading to distinct lifetime and mass evolution. We calculate the resulting modified cosmological constraints and the allowed parameter space to obtain the correct DM abundance. This intriguing scenario provides a unified resolution to both the DM problem and the black hole singularity problem, while preserving the standard self-similar evaporation process.
\end{abstract}
\maketitle

\section{Introduction}

Motivated from the unsatisfactory standard singular black hole (BH) solutions, many attempts have been made to introduce new metrics that could avoid the singularity. Singularity-free black holes are called regular black holes (RBHs). The efforts range from phenomenological metrics such as the Bardeen \cite{Bardeen}, Hayward \cite{Hayward:2005gi}, and Dymnikova \cite{Dymnikova:1992ux} solutions which feature a de Sitter core, or the Culetu-Ghosh-Simpson-Visser metric \cite{Culetu:2013fsa,Culetu:2014lca,Ghosh:2014pba,Simpson:2019mud} which features a Minkowski core, to metrics that are more physically motivated from quantum gravity considerations such as  the Simpson-Visser \cite{Simpson:2018tsi} metric, the Peltola-Kunstatter \cite{Peltola:2008pa,Peltola:2009jm} metric, and the  D’Ambrosio-Rovelli  \cite{Bianchi:2018mml,DAmbrosio:2018wgv} metric. See \cite{Lan:2023cvz,Bambi:2023try} for reviews \footnote{There are also hypothetical horizonless compact objects dubbed black hole mimickers. These objects do not evaporate and are not classified as RBH. See \cite{Bambi:2025wjx} for a review. There is a recent study on possible transitions between different types of objects \cite{Borissova:2025msp}. See also \cite{Rao:2025rop,Huang:2025uhv,Vagnozzi:2022moj,Pedrotti:2024znu} for possible ways to distinguish between regular and singular black holes.}. We note that the stability issue of inner horizon of RBHs is an ongoing research \cite{Carballo-Rubio:2021bpr,Carballo-Rubio:2022kad,Cao:2023par,Bonanno:2022jjp, Bonanno:2022rvo, Bonanno:2025bgc} and some RBH solutions are  actually geodesically incomplete \cite{Zhou:2022yio}. Nevertheless, these  metrics still widely serve as prototype RBH models to explore the resulting rich phenomenologies and demonstrate new research avenues.

Besides the pure theoretical motivations, the observations of too early supermassive black holes \cite{Bogdan:2023ilu,Kovacs:2024zfh} and some intermediate-mass black holes \cite{LIGOScientific:2020iuh,LIGOScientific:2025rsn,IMBHs,Jin:2025izu} have sparked great interest in the idea of cosmologically-coupled black holes \cite{Cadoni:2023lum,Cadoni:2023lqe,Farrah:2023opk,Croker:2024jfg,Faraoni:2024ghi,Calza:2024qxn,Dialektopoulos:2025mfz,Cadoni:2024rri} as traditional mass growth channels \cite{Inayoshi:2019fun,Volonteri:2021sfo} may not be sufficient. These are RBHs that couple with the expansion background such that their mass increases over time that is not due to accretion or mergers. At the current stage, however, the studies are mostly phenomenological and will need stronger theoretical motivations. We do not consider such case here, though it is a potentially interesting direction to consider both cosmological coupling and evaporation effect in a future work.

Meanwhile, primordial black holes (PBHs) are black holes that were produced very early in the Universe \cite{Carr:2009jm}. They were not formed from dead massive stars but from enhanced cosmological perturbations due to, for example, an ultra slow-roll inflationary phase \cite{Ivanov:1994pa,Kinney:2005vj} or a strong first-order phase transition \cite{Hawking:1982ga,Moss:1994pi,Kusenko:2020pcg,Kawana:2021tde}. If these PBHs are also non-singular, we call them regular primordial black holes (RPBHs). Note that the gravitational collapse of usual matter satisfying the standard classical energy conditions would not naturally lead to the formation of RBHs because the Hawking-Penrose theorem dictates that the end point of such process must be a singularity \cite{Penrose:1964wq,Hawking:1970zqf,Hawking:1973uf}. The actual formation of RBHs would require special conditions and, in a recent study \cite{Bueno:2024dgm,Bueno:2024zsx,Bueno:2024eig}, it has been shown that RBHs can exist in $D\geq5$ dimensional spacetime if one has an infinite tower of higher-curvature corrections. However, dynamical formation of RBHs in our usual 4-dimensional spacetime is still a challenging topic but promising progress is being made in this direction \cite{Bueno:2025zaj}. Other models of RBH formation are also being developed \cite{Vertogradov:2025snh,Vertogradov:2025yto,Vertogradov:2025aok,Bonanno:2023rzk,Zhang:2014bea,Mosani:2023awd,Shojai:2022pdq}. Here, we remain agnostic about the RPBHs formation mechanism. Our objective is to study how dark matter (DM) is produced from evaporation of RPBHs given their existence. What we provide is a general framework that could be applied to any static, spherically symmetric RPBH metric that is either phenomenologically or theoretically motivated.

Remnants of non-self-similar RPBHs, such as horizonless compact objects or bare wormholes, could be a DM candidate as studied in \cite{Davies:2024ysj,Trivedi:2025vry,Ong:2024dnr,Dymnikova:2015yma,Bolokhov:2024voa}. Self-similar RPBHs themselves can be a DM candidate in the heavy mass range \cite{Calza:2024fzo,Calza:2024xdh,Calza:2025mwn,Asmanoglu:2025agc}, while particle DM can also be produced from evaporation of ultra-light self-similar RPBHs which is the topic we study here. An advantage of the latter scenario is that particle DM has a much higher chance of being directly detected on earth compared to compact objects like BHs. This is because, for a fixed observed DM energy density, the number density of particle DM must be much larger than that of compact-object DM, significantly increasing the likelihood of interactions with terrestrial detectors. Even without any non-gravitational interaction between DM and ordinary matter, DM may still leave imprints on detectors through their gravitational effects \cite{Badurina:2025xwl,Carney:2019pza}. This choice of scenario is interesting because it simultaneously resolves the DM problem as well as the BH singularity problem, while relying on minimal assumption about the purely gravitational effect of DM \footnote{If one further assumes some form of non-gravitational interaction between DM and ordinary matter, then there may also be other competing production channels from the thermal bath. However, given the current null results from experiments of this kind, we do not make that assumption here.} and the standard semiclassical self-similar BH evaporation. Therefore, in this paper we study particle DM production from evaporation of self-similar RPBHs.

This paper is organized as follows. In Sec. \ref{sec:evaporation of RPBH}, we show how RPBHs evaporate differently from the usual singular PBHs. In Sec. \ref{sec:DM from evaporation of RPBH}, we calculate the DM abundance from evaporation of RPBHs. In Sec. \ref{sec:cosmological constraints}, we point out that some cosmological constraints of this scenario must be modified accordingly. We present our results in Sec. \ref{sec:results}  where the allowed parameter space to obtain the correct DM abundance can be found. Conclusions are in Sec. \ref{sec:conclusion}. Some supplemental details are given in the appendices.  We use the natural units ($\hbar=c=k_\text{B}=1$) in this paper.

\section{Evaporation of regular black holes}\label{sec:evaporation of RPBH}

We consider a class of static, spherically symmetric metrics of the following form:
\begin{equation}\label{eq:metric}
    ds^2=-f(r)dt^2+\frac{dr^2}{g(r)}+h(r)d\Omega^2,
\end{equation}
where $d\Omega^2$ is the usual 2-sphere metric. The metric should be asymptotically flat, meaning that $\{f(r),g(r)\}\xrightarrow{r\rightarrow\infty}1$ and $h(r)\xrightarrow{r\rightarrow\infty}r^2$. As concrete examples, we choose to work with the Hayward metric \cite{Hayward:2005gi} and the Simpson-Visser (SV) metric \cite{Simpson:2018tsi}:
\begin{equation}
    \text{Hayward:}\begin{cases}
        f_\text{Hay}(r)=g_\text{Hay}(r)=1-\frac{2GMr^2}{r^3+2GML^2}\\
        h_\text{Hay}(r)=r^2
    \end{cases},
\end{equation}
\begin{equation}
    \text{Simpson-Visser:}\begin{cases}
        f_\text{SV}(r)=g_\text{SV}(r)=1-\frac{2GM}{\sqrt{r^2+L^2}}\\
        h_\text{SV}(r)=r^2+L^2
    \end{cases}.
\end{equation}
Here, $M>0$ can be interpreted as the BH's mass and $L>0$ is a fundamental length scale known as ``regularizing parameter" which ensures the singularity disappear. One can indeed check that the curvature invariants are finite at $r=0$ (see also \cite{Antonelli:2025zxh}). In the limit $L\rightarrow0$, the above regular metrics reduce to the singular Schwarzschild metric. It can also be seen from these metrics that there exists a maximum value of $L/GM$ above which the horizon does not exist: $L/GM=\sqrt{16/27}$ for Hayward metric and $L/GM=2$ for Simpson-Visser metric.

We chose the Hayward metric as it is well-known as one of the earliest RBH solutions and is usually used as a good benchmark. It could  arise from, for example, theories of nonlinear electrodyanmics \cite{Kumar:2020bqf} or from corrections due to quantum gravity \cite{Addazi:2021xuf}. However, a  more rigorous definition of a singularity-free metric is that it must be geodesically complete, which means that the null/timelike geodesics can be analytically extended beyond $r=0$. Although the curvature invariants of the Hayward metric are everywhere finite, the geodesics are incomplete in this spacetime (at least in the original form) \cite{Zhou:2022yio}. This motivates us to also consider another minimal extension of the Schwarzschild solution which is the Simpson-Visser metric. The curvature invariants of this metric are finite and the geodesics are complete. This metric could emerge in theories of nonlinear electrodynamics with the presence of a phantom scalar field \cite{Bronnikov:2021uta}. Our generic framework can be applied to any other metric of the form given in Eq. \ref{eq:metric}.

\begin{figure}[h!]
    \centering
    \includegraphics[width=0.495\linewidth]{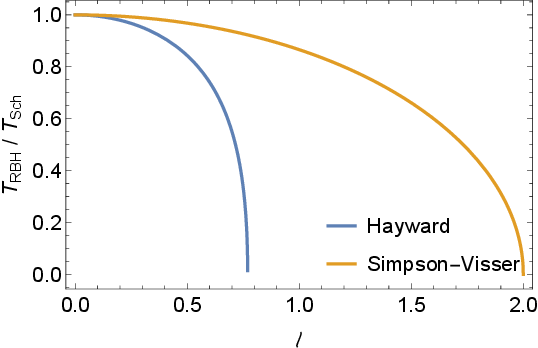}
    \includegraphics[width=0.495\linewidth]{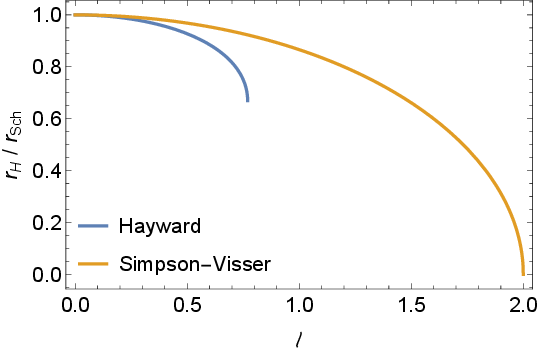}
    \includegraphics[width=0.495\linewidth]{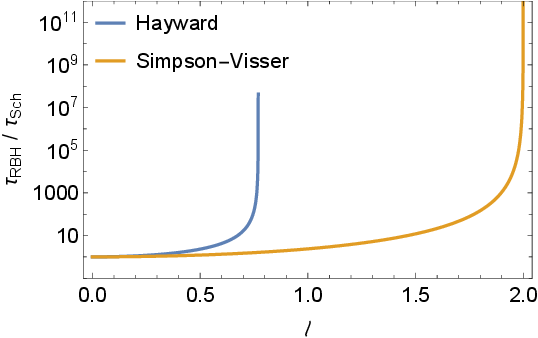}
    \includegraphics[width=0.495\linewidth]{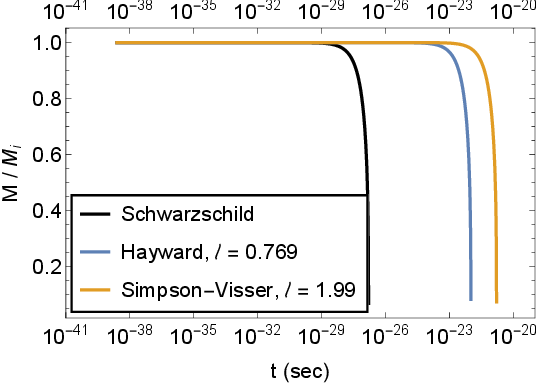}
    \caption{\textit{Upper left:} Hawking temperature of RBH normalized to the Schwarzschild temperature as a function of regularizing parameter $l$. This is $A(l)$ defined in Eq. \ref{eq:A and B}. \textit{Upper right:} Outer horizon radius of RBH normalized to Schwarzschild radius. This is $B(l)$ defined in Eq. \ref{eq:A and B}. \textit{Lower left:} Lifetime of RBH normalized to lifetime of Schwarzschild black hole. \textit{Lower right:} Mass evolution of a 1 g black hole with the regularizing parameters chosen to be close to the extremal values.}
    \label{fig:properties of RBH}
\end{figure}

Just like singular BHs, RBHs also have event horizon and evaporate via Hawking radiation \cite{Hawking:1974rv,Hawking:1975vcx} with the temperature defined as \cite{Gibbons:1977mu}:
\begin{equation}
    T_\text{H}=\frac{\kappa}{2\pi}=\frac{f'(r)}{4\pi}\sqrt{\frac{g(r)}{f(r)}}\Bigg|_{r=r_\text{H}},
\end{equation}
where $\kappa$ is the surface gravity, the prime denotes derivative with respect to $r$, $r_\text{H}$ is the radius of the outer event horizon and is identified as the largest root of the equation $f(r)=0$. As different metrics will lead to different Hawking temperature and horizon size, it is useful to define the ratios of these quantities to that of the Schwarzschild BH:
\begin{equation}\label{eq:A and B}
    A(l)\equiv\frac{T_\text{H}}{T_\text{Sch}}\hspace{1cm},\hspace{1cm}B(l)\equiv\frac{r_\text{H}}{r_\text{Sch}}.
\end{equation}
Here, we defined the dimensionless parameter:
\begin{equation}\label{eq:dimensionless l}
    l\equiv\frac{L}{GM}.
\end{equation}
The quantities with the subscript ``Sch" are associated with the Schwarzschild metric and are given by:
\begin{equation}
    T_\text{Sch}=\frac{1}{8\pi G M}\hspace{1cm},\hspace{1cm}r_\text{Sch}=2GM.
\end{equation}
The dimensionless ratios $A(l)$ and $B(l)$ serve as effective indicators of the underlying physics governing Hawking temperature and horizon size in the subsequent analysis. For the Hayward and Simpson-Visser metrics that we consider, $\{A(l),B(l)\}<1$ (see the upper panels of Fig. \ref{fig:properties of RBH}).

\begin{figure}[h!]
    \centering
    \includegraphics[width=0.8\linewidth]{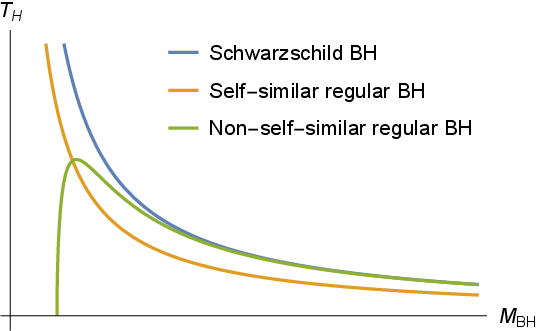}
    \caption{Schematic sketch comparing Hawking temperature of singular Schwarzschild BH with self-similar and non-self-similar types of regular BH. For fixed $l$, we have self-similar singular Schwarzschild BH $(l=0)$ and self-similar RBH $(l>0)$. For time-dependent $l$, we have non-self-similar RBH.}
    \label{fig:TH_sketch}
\end{figure}

RBHs can be classified into self-similar and non-self-similar types as follows (see also Appendix \ref{sec:Appendix self-similar}):
\begin{itemize}
    \item The first approach is that one can fix $L$ and let $M$ vary. This way will lead to stable remnants \cite{Davies:2024ysj} or a transition into a wormhole \cite{Bolokhov:2024voa} because, as $M$ decreases due to evaporation, the ratio $L/GM$ (or $l$ defined in Eq. \ref{eq:dimensionless l}) will increase and eventually reach the extremal value to stop evaporation completely (see the upper left panel in Fig. \ref{fig:properties of RBH} below). However, this approach will also violate the standard self-similarity assumption of BH evaporation, meaning that BH does \textit{not} maintain the semi-classical relations between its parameters throughout the evaporation process (i.e., quantities such as $T_\text{H}M$ and $r_\text{H}/M$ would be time-dependent)\footnote{There is a very nice analogy coming from the concept of fractal geometry in mathematics. In fractal geometry, the geometrical pattern at large scale repeats itself at arbitrarily small scales. This is also known as ``self-similarity".}. We call this non-self-similar RBH. The Hawking temperature of non-self-similar RBH reaches a maximum value and then drops to zero to form remnants (see Fig. \ref{fig:TH_sketch}). We do not follow this approach here. This picture of evaporation was also criticized in \cite{Khodadi:2025icd}.
\item The second approach that we propose in this paper is that we instead fix the dimensionless parameter $l$ defined in Eq. \ref{eq:dimensionless l} \footnote{We think Refs. \cite{Calza:2024fzo,Calza:2024xdh,Calza:2025mwn} also used this approach. While they did not elaborate much on the self-similar nature of RBH evaporation, the fact that they fixed $L/GM$ in their Hawking calculations (or $l$ in our definition) showed that they followed this approach.}. In other words, we redefine the regularizing parameter $L$ such that the factor $1/GM$ is absorbed into its definition. Explicitly, the modified lapse functions are
\begin{equation}
    f_\text{Hay}(r)=1-\frac{2(r/GM)^2}{(r/GM)^3+2l^2},
\end{equation}
\begin{equation}
    f_\text{SV}(r)=1-\frac{2}{\sqrt{(r/GM)^2+l^2}},
\end{equation}
where $0<l<\sqrt{16/27}$ for Hayward metric and $0<l<2$ for Simpson-Visser metric. The geometric properties of this modified version such as the finiteness of curvature invariants, the geodesic completeness, or horizon stability remain intact (see Appendix \ref{sec:Appendix curvature and geodesic}).  In this case, however, the BH will evaporate more slowly but never actually stop evaporation completely until zero mass and there is no remnant. This proposal will preserve the conservative self-similarity assumption of BH evaporation, meaning that BH maintains the semi-classical relations between its parameters throughout the evaporation process (i.e., quantities such as $T_\text{H}M$ and $r_\text{H}/M$ would be constant). This is a standard behavior that we do see in singular BH as well. We call this self-similar RBH. The Hawking temperature of self-similar RBH has a similar behavior to that of the Schwarzschild BH but the scale is adjusted by the choice of the regularizing parameter $l$ (see Fig. \ref{fig:TH_sketch}). We choose this approach in our paper as our goal is to treat DM as particles produced from completely evaporated RPBHs, rather than as remnants of RPBHs. This choice of scenario offers more feasibility for DM direct detection on earth and it eliminates the need to look for exotic horizonless compact objects or wormholes that are typically the end states of non-self-similar RBH. 
\end{itemize}

The maximum possible mass of PBHs at the formation time is the Hubble horizon mass at that time due to causality. Theoretical considerations showed that PBHs indeed have the initial mass of order of the horizon mass \cite{Hawking:1971ei,Carr:1975qj,Kohri:2018qtx}. We assume this is also the case for RPBHs:
\begin{equation}
    M_i=\gamma M_H=\gamma\frac{m_{\text{pl}}^2}{2H_\text{i}},
\end{equation}
where $M_H$ is the Hubble horizon mass, $m_\text{pl}= 1.22\times 10^{19}\text{ GeV}= 2.18\times 10^{-5}\text{ g}$ is the Planck mass, and $H_i$ is the Hubble rate at the formation time. While we choose $\gamma \sim \mathcal{O} (1)$ in the following for concreteness, we retain this factor in the relevant equations to accommodate future scenarios where a more accurate picture of RPBH formation may emerge\footnote{This parameterization is also useful if RPBHs form via a mechanism other than inflationary perturbations such as phase transition. In any case, the mass of RPBH can always be written as a fraction of the Hubble horizon mass. If so, the inflation constraint that we will discuss in Sec. \ref{sec:cosmological constraints} is no longer relevant but all other equations remain valid.}. In a radiation-dominated (RD) Universe, this could be written as
\begin{equation}\label{eq:Mi}
    M_i=\gamma\frac{\sqrt{45}\ m_\text{pl}^3}{4\pi^{3/2}g_{*,i}^{1/2}T_i^2},
\end{equation}
or in a normalized form as
\begin{equation}
    \left(\frac{M_i}{\text{g}}\right)=9.45\times 10^{31}\gamma\left(\frac{106.75}{g_{*,i}}\right)^{1/2}\left(\frac{\text{GeV}}{T_i}\right)^2,
\end{equation}
where $g_{*,i}$ is the number of effective relativistic species at the formation time and $T_i$ is the corresponding plasma temperature. Utilizing the fact that $t=1/2H$ during a RD Universe, it can also be shown that the formation time of RPBH is
\begin{equation}\label{eq:t_i}
    \left(\frac{t_i}{\text{sec}}\right)=2.48 \times 10^{-39}\gamma^{-1}\left(\frac{M_i}{\text{g}}\right).
\end{equation}

Ignoring the greybody factors, BH can be treated as a black body \footnote{In fact, even if considering the full greybody factors, the difference in the results for singular PBHs is less than one order of magnitude \cite{Cheek:2021odj}. It also turns out that the main difference between regular and singular BHs comes from the Hawking temperature, not the greybody factors \cite{Calza:2024fzo}. Therefore, our approximation is good enough to retain both accuracy and analytical insights.}. The energy spectrum of the $i^{th}$ specie emitted from a BH with zero spin and zero charge is:
\begin{equation}
    \frac{d^2u_i(E,t)}{dtdE}=\frac{g_i}{8\pi^2}\frac{E^3}{e^{E/T_H}\pm 1},
\end{equation}
where $u_i(E,t)$ is the total radiated energy per unit area, $g_i$ is the internal degree of freedom of the particle and $E$ is particle's energy. The plus sign is Fermi-Dirac distribution for fermions and the minus sign is the Bose-Einstein distribution for bosons. The mass loss rate is then:
\begin{align}
     \frac{dM}{dt}&=-4\pi r_\text{H}^2\sum_i\int_0^\infty\frac{d^2u_i(E,t)}{dtdE}dE\\
     &=-B(l)^2\frac{2M^2}{\pi m_\text{pl}^4}\sum_ig_i\int_0^\infty\frac{E^3}{e^{E/T_\text{H}}\pm 1}dE\\
     &=-B(l)^2\frac{2M^2}{\pi m_\text{pl}^4}\sum_ig_i\left(\frac{\pi^4T_\text{H}^4}{15}\text{ for bosons, }\frac{7\pi^4T_\text{H}^4}{120}\text{ for fermions}\right)\\
     &=-B(l)^2g_*(T_\text{H})\frac{2\pi^3M^2}{15m_\text{pl}^4}T_\text{H}^4\\
     &=-B(l)^2A(l)^4\frac{g_*(T_\text{H})}{30720\pi}\frac{m_\text{pl}^4}{M^2},
\end{align}
where the appearance of $A(l),B(l)$ comes from Eq. \ref{eq:A and B} and $
    g_*(T_\text{H})=g_B(T_\text{H})+\frac{7}{8}g_F(T_\text{H})$
is the total number of bosonic and fermionic relativistic degrees of freedom at temperature $T_\text{H}$. We can integrate this equation to obtain the mass evolution of the black hole \footnote{Note that $A(l)$ and $B(l)$ are constants for fixed $l$, so they can be pulled out of the integrals.}:
\begin{equation}
    \int_{M_i}^MM^2dM=-B(l)^2A(l)^4\frac{g_*(T_\text{H})m_\text{pl}^4}{30720\pi}\int_{t_i}^tdt
\end{equation}
\begin{equation}
    \Rightarrow M^3-M_i^3=-B(l)^2A(l)^4\frac{g_*(T_\text{H})m_\text{pl}^4}{10240\pi} (t-t_i)
\end{equation}
\begin{equation}\label{eq:M(t)}
    \Rightarrow M(t)=\left(M_i^3-B(l)^2A(l)^4\frac{g_*(T_\text{H})m_\text{pl}^4}{10240\pi}(t-t_i)\right)^{1/3}=M_i\left(1-\frac{t-t_i}{\tau}\right)^{1/3},
\end{equation}
where
\begin{equation}\label{eq:tau_not normalized}
    \tau\equiv\frac{1}{B(l)^2A(l)^4}\frac{10240\pi M_i^3}{g_*(T_\text{H})m_\text{pl}^4}
\end{equation}
is the black hole's lifetime for obvious reason. Putting in numerical factors, we get
\begin{equation}\label{eq:tau}
    \left(\frac{\tau}{\text{sec}}\right)=1.57\times 10^{-27}\frac{1}{B(l)^2A(l)^4}\left(\frac{106.75}{g_*(T_\text{H})}\right) \left(\frac{M_i}{\text{g}}\right)^3.
\end{equation}
From Eqs. \ref{eq:t_i} and \ref{eq:tau}, we see that
\begin{equation}
    \frac{t_i}{\tau}=1.58\times 10^{-12}\gamma^{-1} A(l)^4B(l)^2\left(\frac{g_*(T_\text{H})}{106.75}\right)\left(\frac{\text{g}}{M_i}\right)^2.
\end{equation}
Even for the smallest PBH mass of order 1 g (see Sec. \ref{sec:cosmological constraints}), we see that the formation time is much smaller than the PBH's lifetime, which is even truer for the RPBH metrics that we consider as $\{A(l),B(l)\}<1$, so effectively the lifetime of RPBH can also be used as the cosmic time at which it evaporates.

In Fig. \ref{fig:properties of RBH}, we show the main properties of RBH. This includes the Hawking temperature and horizon radius as a function of regularizing parameter $l$ from Eq. \ref{eq:A and B}, lifetime of RBH from Eq. \ref{eq:tau}, and evolution of RBH's mass from Eq. \ref{eq:M(t)}. We see that RBHs ($l>0$) have lower temperature and smaller horizon size than the singular Schwarzschild BH ($l=0$), so their lifetime is extended \footnote{We note that this is not a universal feature of all RBH metrics existing in the literature. For example, the RBH solution proposed in \cite{Zhang:2024ney} actually has higher temperature than the singular Schwarzschild black hole \cite{Calza:2025mwn}. However, most RBH metrics indeed have lower temperature and our generic formalism can also be extended to any static, spherically symmetric RBH metrics anyway.}. Intuitively, a lower Hawking temperature makes the BH lose mass more slowly, and a smaller horizon size reduces the surface area such that quanta have less ``doors" to exit the event horizon; both of these effects contributed to the extended lifetime. They become either horizonless compact object (Hayward metric) or wormhole (Simpson-Visser metric) and stop evaporation completely if $l$ takes the extremal value: $l=\sqrt{16/27}$ for Hayward metric and $l=2$ for Simpson-Visser metric. Therefore, the close-to-extremal values of $l$ were chosen in the lower right panel of Fig. \ref{fig:properties of RBH} to demonstrate the delayed mass evolution and to highlight the difference from singular Schwarzschild BH.

The cosmic time at which RPBHs evaporate can be estimated from $\Gamma_\text{BH}\sim H_\text{eva}$, where $\Gamma_\text{BH}$ is decay rate of BH, which leads to the corresponding plasma temperature
\begin{equation}\label{eq:T_eva}
    T_\text{eva}\simeq A(l)^2B(l)\frac{\sqrt{3}}{64\times 5^{1/4}}\frac{g_{*}(T_\text{H})^{1/2}m_\text{pl}^{5/2}}{\pi^{5/4}g_{*,\text{eva}}^{1/4}M_i^{3/2}},
\end{equation}
or in the normalized form as
\begin{equation}
    \left(\frac{T_\text{eva}}{\text{GeV}}\right)=1.73\times 10^{10}A(l)^2B(l)\left(\frac{g_{*}(T_\text{H})}{106.75}\right)^{1/2}\left(\frac{106.75}{g_{*,\text{eva}}}\right)^{1/4} \left(\frac{\text{g}}{M_i}\right)^{3/2}.
\end{equation}
The subscript ``eva" denotes the quantities at the evaporation time.

Next, we calculate the number of emitted particles per BH as:
\begin{equation}
    \frac{d^2N_i}{dtdE}=\frac{4\pi r_\text{H}^2}{E}\frac{d^2u_i}{dtdE}=\frac{2g_iB(l)^2}{\pi m_\text{pl}^4}\frac{M^2E^2}{e^{E/T_\text{H}}\pm 1}.
\end{equation}
If the initial Hawking temperature is greater than the particle mass $T_\text{H,in}>m_i$, where $m_i$ is the mass of particle $i^{th}$, evaporation happens throughout the BH's lifetime. Thus, the total number of the $i^{th}$ particle produced is:
\begin{align}
    N_i&=\frac{2g_iB(l)^2}{\pi m_\text{pl}^4}\int_{t_i}^{t_i+\tau}M^2dt\int_0^\infty\frac{E^2dE}{e^{E/T_\text{H}}\pm 1}\\
    &=\frac{2g_i}{\pi m_\text{pl}^4A(l)^4}\frac{30720\pi}{g_*(T_\text{H})m_\text{pl}^4}\int_0^{M_i}M^4dM\int
    _0^\infty\frac{E^2dE}{e^{E/T_\text{H}}\pm 1}\\
    &=\frac{61440g_i}{g_*(T_\text{H})m_\text{pl}^8A(l)^4}\int_0^{M_i}M^4dM\left(2T_\text{H}^3\zeta(3)\text{ for bosons, }\frac{3}{2}T_\text{H}^3\zeta(3)\text{ for fermions}\right),
\end{align}
where $\zeta(3)\approx 1.2$ is the Riemann zeta function at 3. For bosons, we get:
\begin{equation}\label{eq:N_chi for T_H>m_chi}
    N_i=\frac{120g_i\zeta(3)M_i^2}{g_*(T_\text{H})\pi^3m_\text{pl}^2A(l)}.
\end{equation}
If $T_\text{H,in}<m_i$, evaporation happens at the late stage of black hole's evaporation when $T_\text{H}=m_i$, so the number of emitted particles is
\begin{align}
    N_i&=\frac{61440g_i}{g_*(T_\text{H})m_\text{pl}^8A(l)^4}\int_0^{A (l)m_\text{pl}^2/8\pi m_i}M^4dM\int_0^\infty\frac{E^2dE}{e^{E/T_\text{H}}\pm 1}\\
    &=\frac{61440g_i}{g_*(T_\text{H})m_\text{pl}^8A(l)^4}\int_0^{A (l)m_\text{pl}^2/8\pi m_i}M^4dM\left(2T_\text{H}^3\zeta(3)\text{ for bosons, }\frac{3}{2}T_\text{H}^3\zeta(3)\text{ for fermions}\right).
\end{align}
For bosons, we get:
\begin{equation}\label{eq:N_chi for T_H<m_chi}
    N_i=\frac{15g_i\zeta(3)A(l)m_\text{pl}^2}{8g_*(T_\text{H})\pi^5m_i^2}.
\end{equation}
For fermions, we simply multiply the above results by a factor of $3/4$.

\section{Dark matter abundance}\label{sec:DM from evaporation of RPBH}

For the scenario of DM production from evaporation of RPBHs, it is important to determine if there was RPBH domination at early times. As the population of RPBHs dilutes slower than radiation, they may eventually dominate the Universe. In a RD Universe, the fraction of RPBH at the formation time is defined as $\beta\equiv\rho_\text{RPBH,i}/\rho_\text{rad,i}\ll 1$, where $\rho _{\mathrm{RPBH,i}}$ and $\rho _{\mathrm{rad,i}}$ denote the initial energy densities of RPBHs and radiation respectively. Since $\rho_\text{RPBH}/\rho_\text{rad}\propto a\propto T^{-1}$, where $a$ is the scale factor and $T$ is cosmic temperature, it is possible that RPBH may eventually dominate as the Universe expands and temperature drops. However, there will be no RPBH domination if RPBHs evaporate before the matter-radiation equality time \footnote{As RPBHs behave like matter, what we really mean here is the RPBH-radiation equality where RPBHs begin to dominate the Universe, not the subsequent RPBH-radiation equality when RPBHs decay or the late, standard matter-radiation equality at redshift $z\approx 3400$.}: $t_\text{eva}<t_\text{eq}\Rightarrow H_\text{eva}>H_\text{eq}\Rightarrow T_\text{eva}>T_\text{eq}=\beta T_i$. Therefore, the critical population below which there is no RPBH domination is:
\begin{align}
    \beta_c&=\frac{T_\text{eva}}{T_i}\\
    &=\frac{g_{*}(T_\text{H})^{1/2}\ g_{*,i}^{1/4}m_\text{pl}A(l)^2B(l)}{32\sqrt{5\pi\gamma}\ g_{*,\text{eva}}^{1/4}M_i}\\
    &=1.78\times 10^{-6}\gamma^{-1/2} A(l)^2B(l)\left(\frac{g_{*}(T_\text{H})}{106.75}\right)^{1/2}\left(\frac{106.75}{g_{*,\text{eva}}} \right)^{1/4} \left(\frac{g_{*,\text{i}}}{106.75}\right)^{1/4}\left(\frac{\text{g}}{M_i}\right),\label{eq:betac}
\end{align}
where we used Eqs. \ref{eq:Mi} and \ref{eq:T_eva}. The script ``i" denotes values at the formation time of RPBHs. 

Let $\chi$ be a particle DM candidate. Because $\rho_\chi\propto a^{-3}$ and entropy is conserved after RPBH evaporation, the DM abundance today can be calculated as follows:
\begin{equation}
    \Omega_\chi=\frac{\rho_\chi^0}{\rho_\text{crit}^0}=\frac{1}{\rho_\text{crit}^0}\left(\frac{a_\text{eva}}{a_0}\right)^3\rho_\chi^\text{eva}=\frac{1}{\rho_\text{crit}^0}\frac{g_{s,0}T_0^3}{g_{s,\text{eva}}T_\text{eva}^3}\rho_\chi^\text{eva}=\frac{1}{\rho_\text{crit}^0}\frac{g_{s,0}T_0^3}{g_{s,\text{eva}}T_\text{eva}^3}N_\chi m_\chi n_\text{RPBH}^\text{eva},
\end{equation}
where $\rho_\text{crit}^0$ is the critical energy density today, ``0" denotes the values at the present time, the subscript ``s" means entropy degrees of freedom, and $n_\text{RPBH}$ is the number density of RPBH. For monochromatic mass spectrum, $n_\text{RPBH}\propto a^{-3}\Rightarrow n_\text{RPBH}^\text{eva}=n_\text{RPBH}^\text{i}(a_\text{i}/a_\text{eva})^3$, so that
\begin{align}
    \Omega_\chi&=\frac{1}{\rho_\text{crit}^0}\frac{g_{s,0}}{g_{s,\text{eva}}}\left(\frac{T_0}{T_\text{eva}}\right)^3\left(\frac{a_i}{a_\text{eva}}\right)^3N_\chi m_\chi \frac{\beta}{M_i}\rho_\text{rad}^i\\
    &=\frac{4\pi^3}{45H_0^2m_\text{pl}^2}\frac{g_{s,0}\ g_{*,i}}{g_{s,\text{eva}}}\frac{T_0^3T_i^4}{T_\text{eva}^3}\left(\frac{a_i}{a_\text{eva}}\right)^3N_\chi m_\chi\frac{\beta}{M_i},\label{eq:generic Omega_chi}
\end{align}
where we used $\rho_\text{crit}^0=3H_0^2/8\pi G=3H_0^2m_\text{pl}^2/8\pi$ and $\rho_\text{rad}^i=\pi^2g_{*,i}T_i^4/30$. This is the most generic formula for DM abundance and will be used for each case we consider below.

\begin{enumerate}
    \item If $\beta<\beta_c$:\\
In this case, there is no RPBH domination. Entropy is therefore conserved from $T_i$ to $T_\text{eva}$, so Eq. \ref{eq:generic Omega_chi} becomes:
\begin{equation}\label{eq:beta<beta_c generic}
    \Omega_\chi=\frac{4\pi^3}{45H_0^2m_\text{pl}^2}g_{s,0}T_0^3T_iN_\chi m_\chi\frac{\beta}{M_i}.
\end{equation}
There are two subcases:
\begin{itemize}
    \item If $T_\text{H,in}>m_\chi$, we use Eq. \ref{eq:N_chi for T_H>m_chi} for $N_\chi$ and Eq. \ref{eq:Mi}  to obtain:
    \begin{equation}\label{eq:beta<betaC and TH>mDM}
        \Omega_\chi\simeq 6.91\times 10^8 \gamma^{1/2} \frac{1}{A(l)}g_\chi\left(\frac{106.75}{g_*(T_\text{H})}\right)\left(\frac{106.75}{g_{*,i}}\right)^{1/4}\left(\frac{m_\chi}{\text{GeV}}\right)\left(\frac{M_i}{\text{g}}\right)^{1/2}\beta.
    \end{equation}
    \item  If $T_\text{H,in}<m_\chi$, we use Eq. \ref{eq:N_chi for T_H<m_chi} for $N_\chi$ and Eq. \ref{eq:Mi}  to obtain:
    \begin{equation}\label{eq:beta<betaC and TH<mDM}
        \Omega_\chi\simeq 7.74\times 10^{34} \gamma^{1/2} A(l)g_\chi\left(\frac{106.75}{g_*(T_\text{H})}\right)\left(\frac{106.75}{g_{*,i}}\right)^{1/4}\left(\frac{\text{GeV}}{m_\chi}\right)\left(\frac{\text{g}}{M_i}\right)^{3/2}\beta.
    \end{equation}
\end{itemize}
Here, we used $H_0=67\ \rm km/s/Mpc=1.43 \times 10^{-42}\ GeV$, $T_0=2.3\times 10^{-13}\ \rm GeV$, and $g_{\rm s,0}=3.94$.
\item If $\beta\geq\beta_c$:\\
In this case, there is RPBH domination and entropy is not conserved during this period. We instead use the fact that RPBH evaporates when $\Gamma_\text{RPBH}\sim H_\text{eva}$ to have
\begin{align}\label{eq:rho_RPBH^eva}
\rho_\text{RPBH}^\text{eva}&=\rho_\text{RPBH}^\text{i}\left(\frac{a_i}{a_\text{eva}}\right)^3\\
&=\frac{3m_\text{pl}^2H_\text{eva}^2}{8\pi}\\
&\simeq\frac{3m_\text{pl}^2\Gamma_\text{RPBH}^2}{8\pi}\\
&=A(l)^8B(l)^4\frac{3g_*(T_\text{H})^2m_\text{pl}^{10}}{8\times (10240)^2\pi^3M_i^6},
\end{align}    
\begin{equation}
    \Rightarrow \left(\frac{a_i}{a_\text{eva}}\right)^3=A(l)^8B(l)^4\frac{3g_*(T_\text{H})^2m_\text{pl}^{10}}{8\times (10240)^2\pi^3M_i^6\rho_\text{RPBH}^\text{i}}.
\end{equation}
Using this result and $T_\text{eva}$ from Eq. \ref{eq:T_eva}, Eq. \ref{eq:generic Omega_chi} becomes:
\begin{equation}\label{eq:beta>beta_c generic}
    \Omega_\chi\simeq \frac{\pi^{7/4}}{240\sqrt{3}\times 5^{1/4}H_0^2}A(l)^2B(l)g_{s,0} \frac{g_*(T_\text{H})^{1/2}}{g_{*,\text{eva}}^{1/4}} T_0^3m_\text{pl}^{1/2}N_\chi m_\chi \frac{1}{M_i^{5/2}}.
\end{equation}
There are two subcases:
\begin{itemize}
    \item If $T_\text{H,in}>m_\chi$, we use Eq. \ref{eq:N_chi for T_H>m_chi} for $N_\chi$ to get:
    \begin{equation}\label{eq:beta>betaC and TH>mDM}
        \Omega_\chi\simeq 1229 A(l)B(l)g_\chi\left(\frac{106.75}{g_*(T_\text{H})}\right)^{1/2}\left(\frac{106.75}{g_{*,\text{eva}}}\right)^{1/4}\left(\frac{m_\chi}{\text{GeV}}\right)\left(\frac{\text{g}}{M_i}\right)^{1/2}.
    \end{equation}
    \item  If $T_\text{H,in}<m_\chi$, we use Eq. \ref{eq:N_chi for T_H<m_chi} for $N_\chi$ to get:
    \begin{equation}\label{eq:beta>betaC and TH<mDM}
        \Omega_\chi\simeq 1.38\times 10^{29}A(l)^3B(l) g_\chi\left(\frac{106.75}{g_{*,\text{eva}}}\right)^{1/4}\left(\frac{106.75}{g_*(T_\text{H})}\right)^{1/2}\left(\frac{\text{GeV}}{m_\chi}\right)\left(\frac{\text{g}}{M_i}\right)^{5/2}.
    \end{equation}
\end{itemize}
\end{enumerate}

\section{Cosmological constraints}\label{sec:cosmological constraints}
In this section, we show the main cosmological constraints of our scenario.

\subsection{Inflation constraint}
In principle, the minimum mass of RPBH can be the Planck mass of order $10^{-5}\text{ g}$ as it is the quantum gravity limit below which no classical black hole solution is expected to exist. However, if RPBH forms from inflationary perturbations, the observed upper bound on the energy scale of inflation can be translated into a stronger lower bound on the RPBH mass as follows. The power spectrum of scalar perturbation $\mathcal{P}_\zeta$ and tensor perturbation $\mathcal{P}_h$ are defined as \cite{Baumann:2022mni}:
\begin{equation}
    \mathcal{P}_\zeta(k)\equiv A_s\left(\frac{k}{k_*}\right)^{n_s-1}=\frac{1}{8\pi^2}\frac{1}{\varepsilon}\frac{H^2}{M_\text{pl}^2}\Bigg|_{k=aH},
\end{equation}
\begin{equation}
    \mathcal{P}_h(k)\equiv A_t\left(\frac{k}{k_*}\right)^{n_t}=\frac{2}{\pi^2}\frac{H^2}{M_\text{pl}^2}\Bigg|_{k=aH},
\end{equation}
where $\varepsilon$ is the first slow-roll parameter, $M_\text{pl}$ is the reduced Planck mass, $k_*$ is the pivot reference scale, $(A_s,A_t)$ and $(n_s,n_t)$ are the amplitude and tilt of the (scalar,tensor) perturbations respectively. The tensor-to-scalar ratio is defined as $r\equiv A_t/A_s$ which, at the observed CMB scale, leads to:
\begin{equation}
    r\Bigg|_{k=k_*}=\frac{2}{\pi^2}\frac{H_\text{inf}^2}{M_\text{pl}^2}\frac{1}{A_s}\Bigg|_{k=k_*}.
\end{equation}
At $k_*=0.05\text{ Mpc}^{-1}$, we have $r<0.1$ and $A_s\simeq 2.1\times 10^{-9}$  \cite{Planck:2018jri}. Using $M_\text{pl}\simeq 2.43\times 10^{18}$ GeV, the upper bound on the energy scale of inflation is
\begin{equation}
    \left(\frac{H_\text{inf}}{\text{GeV}}\right)<7.82\times10^{13}.
\end{equation}
By using the standard Friedmann equation for a RD Universe $H=\sqrt{\pi^2g_*(T_{\text{reh}})T_{\text{reh}}^4/90M_{\text{pl}}^2}$, assuming instant reheating, this can also be given in terms of the reheating temperature as:
\begin{equation}
    \left(\frac{T_\text{reh}}{\text{GeV}}\right)<7.45\times 10^{15},
\end{equation}
where $g_*(T_{\text{reh}})\sim 106.75$ has been used. Using this result in Eq. \ref{eq:Mi} then gives a lower bound on the RPBH mass:
\begin{equation}\label{eq:inflation constraint}
    \left(\frac{M_i}{\text{g}}\right)>1.7\ \gamma
\end{equation}

\subsection{BBN constraint}
Big Bang Nucleosynthesis  (BBN) is the moment when temperature is low enough for the formation of light elements' nuclei. This happens at around 1 second or $T\sim O(1\text{ MeV})$ \cite{Baumann:2022mni}. From the observation of light elements' abundances, we know very well the amount of radiation available during this period. Any excessive radiation produced from  RPBHs' evaporation is not allowed. We therefore require that RPBHs must evaporate before BBN in order to be consistent with observations \footnote{Actually, PBH can evaporate during or even after BBN as long as their population is very small \cite{Keith:2020jww}. Such fine-tuned population seems artificial and because we conservatively assume that DM is produced prior to BBN, we do not consider that case.}. From Eq. \ref{eq:tau}, the condition $\tau_\text{RPBH}<1\text{ sec}$ gives an upper bound on the RPBH mass:
\begin{equation}\label{eq:BBN constraint}
    \left(\frac{M_i}{\text{g}}\right)<8.6\times 10^8A(l)^{4/3}B(l)^{2/3}.
\end{equation}

\begin{figure}[h!]
    \centering
    \includegraphics[width=0.8\linewidth]{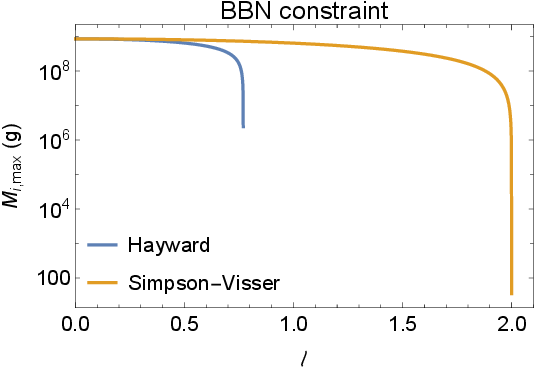}
    \caption{BBN constraint on RPBH mass as a function of $l$.}
    \label{fig:BBN constraint}
\end{figure}

In Fig. \ref{fig:BBN constraint}, we show the allowed maximum RPBH mass from BBN constraint as a function of $l$. For the singular Schwarzschild metric, PBHs with mass below $10^9\text{ g}$ are allowed. For RPBHs, however, the constraint is stronger because RPBHs live longer and may evaporate too late.

\subsection{Warm dark matter constraint}
DM produced from evaporation of RPBHs may have large kinetic energy (warm DM) that could erase the observed small-scale structures such as Lyman-$\alpha$ forest or Milky Way satellite. Therefore, we need to put constraint on RPBH such that the DM produced has enough time to lose its momentum due to cosmological redshift. The warm DM constraint of singular PBH type has been studied in the literature such as \cite{Baldes:2020nuv,Masina:2020xhk}. We redo some steps here to take into account modified evaporation dynamics of RPBH. The momentum of DM today is
\begin{equation}\label{eq:p_0}
    p_0=m_\chi v_0=p_\text{eva}\left(\frac{a_\text{eva}}{a_0}\right)\simeq T_\text{H,in}\left(\frac{a_\text{eva}}{a_0}\right)=T_\text{H,in}\left(\frac{g_{s,0}}{g_{s,\text{eva}}}\right)^{1/3}\left(\frac{T_0}{T_\text{eva}}\right),
\end{equation}
where $v_0$ is the velocity today and the average initial momentum of evaporated DM was taken to be the initial Hawking temperature of RPBH \footnote{This is a reasonable approximation as the warm DM constraint is only important for light DM mass such that $T_\text{H,in}>m_\chi$. As RPBH mass only changes significantly at the end of its lifetime, most DM produced should have initial kinetic energy (or momentum) of order $T_\text{H,in}$.}. The smallest scale that could be probed is of order $k\sim 10\text{ Mpc}^{-1}$, which corresponds to the photon temperature of order $T_\gamma\sim 1\text{ keV}$ \cite{Lin:2019uvt}. Therefore, thermally produced warm DM must be heavier than 1 keV in order to be non-relativistic by that time. Indeed, it was found in \cite{Villasenor:2022aiy} that thermally produced warm DM must be heavier than 3.5 keV. As entropy is conserved from the decoupling time to now, this lower bound on DM mass can be translated into a universal upper bound on current DM velocity for a generic warm DM candidate \cite{Masina:2020xhk}:
\begin{equation}
    v_0=a_\text{dec}=\left( \frac{g_{s,0}}{g_{s,\text{dec}}}\right)^{1/3}\left(\frac{T_0}{T_\text{dec}}\right)=\left( \frac{g_{s,0}}{g_{s,\text{dec}}}\right)^{1/3}\left(\frac{T_0}{m_\chi }\right),
\end{equation}
where $g_{s,0}=3.94$ and $g_{s,\text{dec}}\sim 106.75$ have been used. So that $m_\chi>3.5\text{ keV}$ gives $v_0<2.19\times 10^{-8}$. By using Eqs. \ref{eq:p_0}, \ref{eq:A and B}, and \ref{eq:T_eva}, we obtain an upper bound on RPBH mass:
\begin{equation}\label{eq:WDM constraint}
    \left(\frac{M_i}{\text{g}}\right)\lesssim 2.18\times 10^5A(l)^2B(l)^2\left(\frac{m_\chi}{\text{GeV}}\right)^2.
\end{equation}

\begin{figure}[h!]
    \centering
    \includegraphics[width=\linewidth]{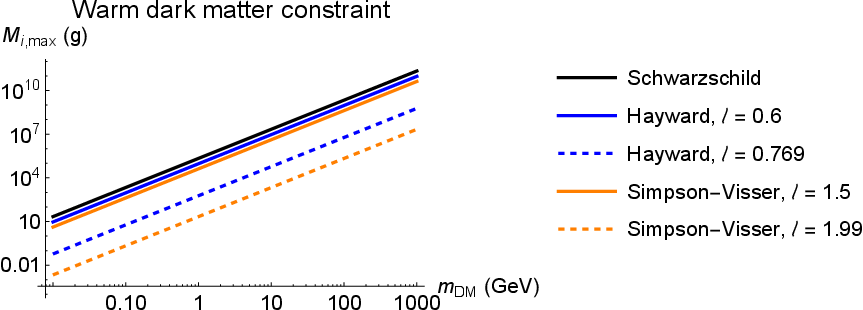}
    \caption{Warm DM constraint on RPBH mass as a function of DM mass for different chosen values of $l$.}
    \label{fig:WDM constraint}
\end{figure}

In Fig. \ref{fig:WDM constraint}, we show the warm DM constraint on the RPBH mass as a function of DM mass for different chosen values of $l$. We see that larger DM mass allows larger PBH mass as the initial kinetic energy of DM is smaller and therefore the small-scale structure constraint becomes more irrelevant. For singular Schwarzschild PBH, the entire parameter space is allowed if $m_\chi\gtrsim 100\text{ GeV}$ (recall that $M_i<10^9\text{ g}$ from BBN constraint). For RPBHs, the constraint is shifted towards lower mass because the RPBHs that we consider have lower Hawking temperature and smaller horizon size leading to longer lifetime (Eq. \ref{eq:tau}), which means that DM could be produced too late and does not have enough time to become cold enough by the time small-scale structures form. An increase of $l$ results in stronger constraints due to the more extended lifetime of the RPBH (see the lower left panel of Fig. \ref{fig:properties of RBH}). Therefore, heavy DM is generally preferred in this scenario.


\section{Results}\label{sec:results}
\begin{figure}[h!]
    \centering
    \includegraphics[width=0.57\linewidth]{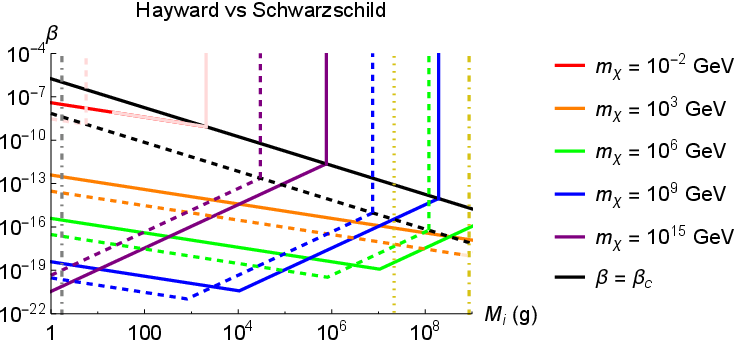}
    \includegraphics[width=0.42\linewidth]{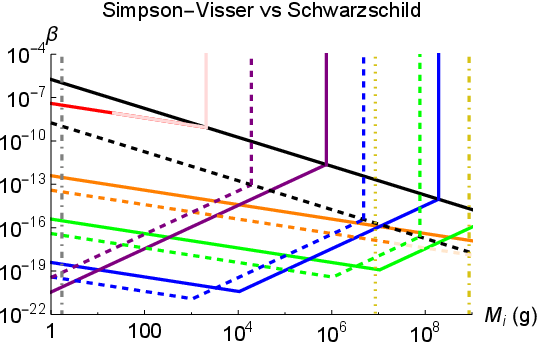}
    \caption{\textit{In the left panel}, the solid colored contours show the parameters that can obtain the correct DM abundance for the singular Schwarzschild PBH with different DM masses. The solid black line indicates $\beta=\beta_c$ above which there is PBH domination. The dashed colored and black contours are for the regular Hayward metric with $l=0.769$. The vertical dot-dashed gray line is inflation constraint for both PBH and RPBH. The vertical dot-dashed yellow line is BBN constraint for PBH, while the dotted yellow line is for RPBH. The faded portions of (solid or dashed) colored contours, if any, do not satisfy warm DM constraint. \textit{In the right panel}, the same but for Simpson-Visser metric with $l=1.99$.}
    \label{fig:DM abundance}
\end{figure}

We now show in Fig. \ref{fig:DM abundance} the allowed parameter space on the population-mass plane of RPBH where the observed DM abundance can be achieved. In order to highlight the difference between regular and singular PBHs, the regularizing parameters are chosen to be close to the extremal values: $l=0.769$ for Hayward metric (left panel) and $l=1.99$ for Simpson-Visser metric (right panel). For comparison, result of the usual singular Schwarzschild PBH case \cite{Gondolo:2020uqv,Cheek:2021odj,Lennon:2017tqq} is also shown which can be easily obtained from our above formulae by taking the limit $l\rightarrow 0$ corresponding to $\{A(l),B(l)\}\rightarrow 1$. We consider a scalar DM particle with $g_\chi=1$.

The solid colored contours are parameters of singular Schwarzschild PBH that can achieve the observed DM abundance $\Omega_\chi\simeq 0.27$ \cite{Planck:2018vyg}. The region above and to the left of the colored contours corresponds to overproduction of DM, while the region below and to the right indicates underproduction. The region above the solid black line $\beta=\beta_c$ is where there is PBH domination. The corresponding dashed colored and black contours are for the RPBH metrics. The region to the right of the vertical dot-dashed gray line is allowed from inflation constraint for both PBH and RPBH. The region to the left of the vertical dot-dashed yellow line is allowed from the BBN constraint for PBH, whereas the dotted yellow line is for RPBH. The faded portions of the (solid or dashed) colored  contours, if any, are not allowed by the warm DM constraint. Note that for $m_\chi=10^{-2}\text{ GeV}$, the dashed red contour of the Simpson-Visser metric is shifted toward much smaller mass that is below 1 gram and is therefore not visible in the figure.

We first explain the physics behind the parameter space for singular Schwarzschild PBH \cite{Cheek:2021odj,Hooper:2019gtx,Masina:2020xhk,Fujita:2014hha,Gondolo:2020uqv,Morrison:2018xla}:
\begin{itemize}
    \item The region below the critical line $\beta=\beta_c$ is where there is no PBH domination because the population is too small. Heavy PBHs live longer so that the required population to have PBH domination gets smaller (see Eq. \ref{eq:betac}).
    \item The decreasing portion of the solid colored contours describes the case when the initial PBH mass is small enough such that its initial Hawking temperature is greater than DM mass: $T_\text{H,in}>m_\chi$  (see Eq. \ref{eq:beta<betaC and TH>mDM}). In this case, DM production occurs throughout the PBH's lifetime. More PBH mass leads to more number of DM particles produced per PBH (see Eq. \ref{eq:N_chi for T_H>m_chi}). Thus, PBH population must be reduced to compensate to obtain a fixed observed DM abundance.
    \item When the PBH mass is large enough such that $T_\text{H,in}<m_\chi$, the solid colored contours begin to have an increasing behavior (see Eq. \ref{eq:beta<betaC and TH<mDM}). In this case, DM production occurs at the late stage of PBH evaporation. The number of DM particles produced is independent from PBH mass (see Eq. \ref{eq:N_chi for T_H<m_chi}). The DM abundance is instead  controlled by PBH number density (see Eq. \ref{eq:beta<beta_c generic}). Thus, increasing PBH mass needs increasing PBH energy density to get the same DM abundance.
    \item If $m_\chi$ is too small (large), only the case of $T_\text{H,in}>m_\chi$ ($T_\text{H,in}<m_\chi$) may exist in the allowed PBH mass range. In particular, in the shown PBH mass range from 1 gram to $10^9$ gram, $m_\chi<10^4\text{ GeV}$ ($m_\chi>10^{13}\text{ GeV}$) only has the case of $T_\text{H,in}>m_\chi$ ($T_\text{H,in}<m_\chi$). Thus, in Fig. \ref{fig:DM abundance}, the cases of $m_\chi=10^{-2}\text{ GeV}$ and $m_\chi=10^3\text{ GeV}$ only have the decreasing colored contours corresponding to $T_\text{H,in}>m_\chi$, whereas the case of $m_\chi=10^{15}\text{ GeV}$ only has the increasing colored contour corresponding to $T_\text{H,in}<m_\chi$.
    \item The region above the solid black line is where the population is large enough to have a PBH domination period. The colored contours begin to turn vertical in this region as the DM abundance is independent from PBH population (see Eq. \ref{eq:beta>beta_c generic}). The origin of this independence comes from the fact that PBHs dominated the Universe such that their energy density at the evaporation time controls the Hubble rate at that time, which in turn can be gauged by the decay rate of PBHs that is solely determined by PBHs' mass (see Eq. \ref{eq:rho_RPBH^eva} and follow-up equations).
\end{itemize}

For RPBHs, the main physics is the same but the parameter space is shifted. The crucial point to understand these shifts is that RPBHs have longer lifetime than singular PBHs due to the modification of Hawking temperature and horizon size (Eq. \ref{eq:tau}). With that in mind, we can explain the shifts as follows. The critical line $\beta=\beta_c$ is shifted toward smaller population because, for a given mass, RPBH lives longer than singular PBH so that the population must be smaller in order to not have RPBH domination. Similarly, the decreasing portion of the colored contour corresponding to $\beta<\beta_c$ and $T_\text{H,in}>m_\chi$ is shifted toward smaller population because RPBHs live longer to produce more DM particles per BH (Eq. \ref{eq:N_chi for T_H>m_chi}), so that the population must be smaller to compensate (Eq. \ref{eq:beta<beta_c generic}). As a small-mass RPBH would have a similar temperature to a large-mass singular PBH, the turning point at which $T_\text{H,in}=m_\chi$ must be shifted toward smaller RPBH mass. The increasing portion of colored contours corresponding to $\beta<\beta_c$ and $T_\text{H,in}<m_\chi$ is shifted toward larger population because RPBHs have lower temperature and produce less number of DM particles per BH (Eq. \ref{eq:N_chi for T_H<m_chi}), so that their population must be enhanced to compensate (Eq. \ref{eq:beta<beta_c generic}). Finally, the vertical portion of the colored contours corresponding to $\beta>\beta_c$ is shifted toward smaller mass as RPBHs' lifetime is prolonged. Overall, we observe that the parameter space shifts more significantly in the region with PBH domination ($\beta>\beta_c$) than in the region without it ($\beta<\beta_c$). This fact can also be seen from Eq. \ref{eq:beta>beta_c generic} ($\beta>\beta_c$) which has stronger dependence on  RBHs' parameters $A(l)$ and $B(l)$ than Eq. \ref{eq:beta<beta_c generic} ($\beta<\beta_c$). The shifts of BBN and warm DM constraints were already discussed in Sec. \ref{sec:cosmological constraints}.

From the above discussion, we can see the power of our analytical treatment where, at every step of calculation, the relevant physics reveals itself in a completely transparent way. The dimensionless ratios $A(l)$ and $B(l)$ can be used to indicate exactly whether the modifications of physical quantities are from modified Hawking temperature or modified horizon radius respectively, and how much they contribute. Our generic machinery are valuable for model builders to have preliminary assessments of model's observational features.

Importantly, we see that the allowed parameter space for RPBH could be shifted by many orders of magnitude compared to the singular PBH case, especially if the regularizing parameter $l$ is close to extremal. Some previously allowed regions are now ruled out if BH is regular. This observation reveals that BH's interior, while hidden inside the event horizon, can actually reveal itself via the evaporation product and can leave distinctive observational imprints. For instance, for a fixed DM mass, a two-order-of-magnitude change in the initial BH mass can lead to a one-order-of-magnitude shift in the gravitational wave (GW) frequency associated with PBH formation, since $f_{\mathrm{GW}}\propto M^{-1/2}$. This simple example hints at a whole new avenue to use multiple observational channels to probe and distinguish between singular and regular BH.

\section{Conclusion}\label{sec:conclusion}
In this paper, we showed that a simple redefinition of the regularizing parameter in RBH metric can preserve the self-similarity of the evaporation process. This realization implies that BH can still be regular without ending up in exotic, unverified remnant states such as horizonless compact object or wormhole. Assuming the existence of RPBHs in the early Universe and the particle nature of DM, we then presented a scenario in which particle DM is produced from evaporation of self-similar RPBHs. This proposal is compelling because the particle nature of DM enhances the feasibility of terrestrial direct detection experiments compared to compact-object DM, and it also resolves the BH singularity problem while maintaining the standard self-similar evaporation process. We demonstrated the idea with some benchmark RBH solutions including the Hayward and the Simpson-Visser metrics to show how the resulting allowed parameter space is modified. Our general framework can be applied to any other static, spherically symmetric RPBH metric by using the parameterization given in Eq. \ref{eq:A and B}, along with the analytical formulae for DM abundance in Eqs. \ref{eq:beta<betaC and TH>mDM}, \ref{eq:beta<betaC and TH<mDM}, \ref{eq:beta>betaC and TH>mDM}, and \ref{eq:beta>betaC and TH<mDM} as well as the cosmological constraints in Eqs. \ref{eq:inflation constraint}, \ref{eq:BBN constraint}, and \ref{eq:WDM constraint}. The parameter space could be constrained further by using $\Delta N_\text{eff}$ constraint on isocurvature GW when there is RPBH domination \cite{Loc:2025whx}. Our work offered ready-to-use analytical formulae and transparent physical insights which could be tracked at any stage of the calculation. For example, the prolonged lifetime of RPBH in Eq. \ref{eq:tau} has contribution from both a smaller horizon size ($B(l)<1$) and a lower Hawking temperature ($A(l)<1$). BH model builders could easily utilize our results to visualize the resulting parameter space of their model and to keep track of the model's parameter.

We note that this work serves as a pilot study, opening up a wide array of compelling directions for future investigations. First, we assumed that the initial mass of RPBH is
the same as that of singular PBH. The linear dependence of initial PBH mass on Hubble horizon mass is likely a robust expectation as the latter sets the mass scale for various physical phenomena in the early Universe, while the proportional constant $\gamma$ captures the details of particular models. We chose this factor to be $\sim\mathcal{O}(1)$ as we tried to be agnostic about specific formation models. A more dedicated model-dependent study regarding the formation of RPBH is needed to specify more accurately the proportional constant $\gamma$, which could affect some of the results presented here. Our equations keep track of this parameter so that the results could be updated conveniently and one can easily see the levels of dependence on this constant. Second, while both the original chosen RBH metrics and the modified versions introduced in our paper were constructed mainly based on desirable phenomenological features, it is interesting to explore possible realization of such scenarios from a more theoretically motivated viewpoint. Our framework applies to any static, spherically symmetric RBH model which admits self-similar evaporation. Third, it is known that the emission spectra is most sensitive to the modified Hawking temperature and not the modified greybody factors \cite{Calza:2024fzo}, so it was a good approximation to utilize the blackbody approximation in this paper to provide analytical insights and to keep track of the physics involved when comparing with the singular PBH case. An extension of our work could be to include the greybody factors to obtain more refined accuracy once a specific, well-motivated metric is chosen. Fourth, we conservatively considered the standard radiation dominated Universe for RPBH formation which is expected to lead to negligible BH's spin \cite{Chiba:2017rvs,DeLuca:2019buf}. A potential direction is to consider RPBH with high spin which may be possible if they form in a nonstandard thermal history such as early matter domination \cite{Harada:2017fjm}. Due to angular momentum conservation, spinning RPBH could alter the allowed parameter space for high-spin DM particles. Fifth, as a starting point, we assumed an almost monochromatic mass function. Specific models of RPBH formation could lead to specialized forms of extended mass function that deserve further exploration. Finally, as self-similar ultra-light RPBHs no longer exist now, the GW backgrounds associated with their formation, merger, or evaporation could be our best hope to probe them. As we showed in this paper, the allowed parameter space for the scenario of DM production from evaporation of RPBHs could be shifted by many orders of magnitude compared to the singular PBH case. A detailed study of GW therefore offers a compelling way to simultaneously resolve the DM problem and observationally discriminate regular and singular black holes. These exciting topics are left for future works.

\section*{Acknowledgments}
This work was supported by the National Science and Technology Council, the Ministry of Education (Higher Education Sprout Project NTU-114L104022-1), and the National Center for Theoretical Sciences of Taiwan.


\appendix

\section{Self-similar vs non-self-similar regular black hole}\label{sec:Appendix self-similar}
For a clearer comparison between self-similar and non-self-similar types of RBH, let's consider explicitly the Simpson-Visser metric which admits the following form of Hawking temperature and horizon radius \cite{Simpson:2018tsi}:
\begin{equation}
    T_\text{H}=T_\text{Sch}\sqrt{1-\left(\frac{L}{2GM}\right)^2}=\frac{1}{8\pi G M}\sqrt{1-\left(\frac{L}{2GM}\right)^2}=\frac{1}{8\pi GM}\sqrt{1-\left(\frac{l}{2}\right)^2},
\end{equation}
\begin{equation}
    r_\text{H}=r_\text{Sch}\sqrt{1-\left(\frac{L}{2GM}\right)^2}=2GM\sqrt{1-\left(\frac{L}{2GM}\right)^2}=2GM\sqrt{1-\left(\frac{l}{2}\right)^2}.
\end{equation}
\begin{itemize}
    \item If we fix $L$, the quantities $T_\text{H}M$ and $r_\text{H}/M$ would be time-dependent as $M$ decreases due to evaporation. In other words, the BH does \textit{not} maintain the semi-classical relations between its parameters throughout the evaporation process. The Hawking temperature reaches a maximum at $T_\text{H}=1/8\pi L$ for $M=L/\sqrt{2}G$ and then drops to zero at $M=L/2G$. Thus, the end state of such RBH is a stable remnant (a wormhole for the Simpson-Visser metric) with nonzero mass.  This is classified as non-self-similar RBH.
    \item If we fix the dimensionless parameter defined as $l\equiv L/GM$, the quantities $T_\text{H}M$ and $r_\text{H}/M$ would be constants. In other words, the BH maintains the semi-classical relations between its parameters throughout the evaporation process. The BH evaporates in a runaway manner as the Hawking temperature increases when the BH mass decreases. The final state is that the BH disappears completely and there is no remnant. This is classified as self-similar RBH. If we simply set $l=0$, we recover the standard self-similar singular BH with $T_\text{H}M=1/8\pi G$ and $r_\text{H}/M=2G$.
\end{itemize}
A similar analysis for the Hayward metric can be done numerically as shown in the upper panels of Fig. \ref{fig:properties of RBH}. For a fixed $l$, the ratios $T_\text{H}/T_\text{Sch}$ and $r_\text{H}/r_\text{Sch}$ are fixed, so that $T_\text{H}M$ and $r_\text{H}/M$ are constants.

\section{Curvature invariant, geodesic completeness, and horizon stability}\label{sec:Appendix curvature and geodesic}
Here, we show that the utilization of the dimensionless regularizing parameter $l$ preserves the finiteness of curvature invariant and the completeness of the geodesic. The most important curvature invariant to assess whether or not the metric admits singularity is the Kretschmann scalar which is defined as $\mathcal{K}\equiv R_{\mu\nu\rho\sigma}R^{\mu\nu\rho\sigma}$, where $R_{\mu\nu\rho\sigma}$ is the Riemann tensor. A Taylor expansion near $r=0$ of the Hayward and Simpson-Visser (SV) metrics gives \cite{DeLorenzo:2014pta,Simpson:2018tsi}:
\begin{equation}
\text{Hayward:}\hspace{1cm}
    \mathcal{K}^2\xrightarrow{r\rightarrow0}\frac{24}{L^4}+O(r^3)=\frac{24}{G^4M^4l^4}+O(r^3),
\end{equation}
\begin{equation}
    \text{SV: }
    \mathcal{K}^2\xrightarrow{r\rightarrow 0}\frac{4}{L^4}\left(3-\frac{8GM}{L}+\frac{9G^2M^2}{L^2}\right)+O(r^2)=\frac{4}{G^4M^4l^4}\left(3-\frac{8}{l}+\frac{9}{l^2}\right)+O(r^2).
    \end{equation}
As long as $L>0$, the Kretschmann scalar remains finite at $r=0$ for all $M$. In the languague of $l$, the Kretschmann scalar remains finite at $r=0$ for all $M\neq 0$ as long as $l>0$. It only diverges when $M=0$ so the black hole is regular at any moment in its lifetime \footnote{In practice, the minimum mass of BH should be the Planck mass of order $10^{-5}\ \rm g$. Beyond that, it will emit a burst of radiation with energy of order $10^{19}\ \rm GeV$ and disappear, so $M=0$ cannot be reached in reality.}.

Following \cite{Zhou:2022yio}, the radial geodesic of a test particle moving in the metric given in Eq. \ref{eq:metric} is:
\begin{equation}
    \dot{r}^2=e^2-f(r),
\end{equation}
where the dot is derivative with respect to proper time and $e\equiv -g_{tt}\dot{t}$ is the conserved energy associated with the timelike Killing vector of our static metrics. We then have
\begin{equation}
    \text{Hayward:}\hspace{1cm}
    \dot{r}^2=e^2-1+\frac{2GMr^2}{r^3+L^3}=e^2-1+\frac{2(r/GM)^2}{(r/GM)^3+l^3},
\end{equation}
\begin{equation}
    \text{Simpson-Visser:}\hspace{1cm}
    \dot{r}^2=e^2-1+\frac{2GM}{\sqrt{r^2+L^2}}=e^2-1+\frac{2}{\sqrt{(r/GM)^2+l^2}}.
\end{equation}
We see that the geodesic of Hayward metric diverges at $r=-L$, or equivalently at $r=-lGM$ in our redefinition. The geodesic of Simpson-Visser metric is regular for all $r$ in either definition of the regularizing parameter.

Lastly, we note that some metrics such as the Hayward metric has an issue known as mass-inflation instability. This means that under perturbations the mass parameter in a given region of spacetime undergoes exponential amplification within a finite time scale determined by the surface gravity $\kappa_{-}$ of the inner horizon: $|\kappa_-|^{-1}\sim L=GMl$ \cite{Carballo-Rubio:2018pmi}. In either case of fixing $L$ or $l$, this timescale is always smaller than the BH lifetime which makes the BH become inevitably unstable \footnote{For fixed $L$, the BH lifetime is infinite in the sense that it will no longer evaporate once it has transited into some stable remnant state. For fixed $l$, the RBH evaporates completely and its finite lifetime is given in our Eq. \ref{eq:tau_not normalized}.}. Such issue is not present for Simpson-Visser BH as it only has one horizon in our Universe by construction \cite{Simpson:2018tsi}. We note in passing that many RBH models with stable horizons are also being developed recently such as \cite{Carballo-Rubio:2022kad,Cao:2023par,Eichhorn:2025pgy}. While working with some classic RBH models to illustrate our core idea, better future RBH models could utilize our general framework as long as the metric takes the functional form in Eq. \ref{eq:metric} and the Hawking evaporation is self-similar.


\bibliographystyle{JHEP}
\bibliography{references}

\end{document}